\documentclass[aip,jcp,reprint]{revtex4-1}
\usepackage{graphicx,multirow}

\usepackage{epsfig}
\usepackage{graphicx}
\usepackage{amsmath}
\usepackage{amssymb}
\usepackage{amsfonts}  
\usepackage{fancyhdr}
\usepackage{bm}
\usepackage{dcolumn}
\usepackage{subfigure}

\long\def\symbolfootnote[#1]#2{\begingroup%
\def\thefootnote{\fnsymbol{footnote}}\footnote[#1]{#2}\endgroup}

\newcommand{\bra}[1]{\langle #1|}
\newcommand{\ket}[1]{|#1\rangle}

\begin{document}

\title{Density functional theory embedding for correlated wavefunctions: Improved methods for open-shell systems and transition metal complexes} 

\author{Jason D. Goodpaster$^{\dag}$}
\author{Taylor A. Barnes$^{\dag}$}
\author{Frederick R. Manby$^{\ddag}$}
\author{Thomas F. Miller, III$^{\dag}$}
\email{tfm@caltech.edu}

\affiliation{
$^{\dag}$Division of Chemistry and Chemical Engineering, California Institute of Technology, Pasadena, CA 91125, USA \\
$^\ddag$Centre for Computational Chemistry, School of Chemistry, University of Bristol, Bristol BS8 ITS, United Kingdom
}

\date{\today}

\begin{abstract}
Density functional theory (DFT) embedding provides a formally exact framework for interfacing correlated wave-function theory (WFT) methods with lower-level descriptions of electronic structure.  Here, we report techniques to improve the accuracy and stability of  WFT-in-DFT embedding calculations.  In particular, we develop spin-dependent embedding potentials in both restricted and unrestricted orbital formulations to enable WFT-in-DFT embedding for open-shell systems, and we develop an orbital-occupation-freezing technique to improve the convergence of optimized effective potential (OEP) calculations that arise in the evaluation of the embedding potential.  The new techniques are demonstrated in applications to the van-der-Waals-bound ethylene-propylene dimer and to the hexaaquairon(II) transition-metal cation.  Calculation of the dissociation curve for the ethylene-propylene dimer reveals that WFT-in-DFT embedding reproduces full CCSD(T) energies to within 0.1 kcal/mol at all distances, eliminating  errors in the dispersion interactions due to conventional exchange-correlation (XC) functionals while simultaneously avoiding errors due to subsystem partitioning across covalent bonds.  Application of WFT-in-DFT embedding to the calculation of the low-spin/high-spin splitting energy in the hexaaquairon(II) cation reveals that the majority of the dependence on the DFT XC functional can be eliminated by treating only the single transition-metal atom at the WFT level; furthermore, these calculations demonstrate the substantial effects of open-shell contributions to the embedding potential, and they suggest that restricted open-shell WFT-in-DFT embedding provides better accuracy than unrestricted open-shell WFT-in-DFT embedding due to the removal of spin contamination.
 
\end{abstract}

\maketitle

\section{Introduction}

The demand for accurate and efficient descriptions of complex molecular systems requires development of quantum embedding methods for electronic structure in which a small subsystem is treated with a high level of theory while the remainder of the system is treated at a more affordable level.
Widely used examples of quantum embedding include QM/MM,\cite{Levitt76,Hillier97,Field1998,Truhlar07,Thiel09,Ryde11} ONIOM,\cite{Frisch99,Mor95} and fragment molecular orbital (FMO) approaches,\cite{Nakano99,Kitaura2004,Kitaura2007} which have led to significant advances in the simulation 
of condensed-phase and biomolecular systems.  However, such methods generally rely on empirical models for the subsystem interactions, including link-atom approximations for embedding across covalent bonds\cite{Kollman86,Karplus90,Yang99,Zhang05} and point-charge electrostatic descriptions of the environment,\cite{Truhlar07,Nakano99} that are difficult to systematically improve and that can fail in practical applications.\cite{Field1998,Ryde11,Scheiner91,Langell2007}

Density functional theory (DFT) offers an appealing framework for addressing this challenge.\cite{Subbaswamy86,Subbaswamy87,Cor91,wesorig,Wes06,carterembedding,Car11b,Was10,Was11,tfm10jason,tfm11jason,tfm12jason,Rei08,Vis10b,Ron08,Ron09,Johannes12,Jac07,Ian06,Kam10,Vis09,Vis08b,Car11a,Cha10,Jacob12}  
DFT embedding provides a formulation of electronic structure theory in which subsystem interactions 
depend only on their electronic densities, including  
 non-additive contributions due to the electrostatic, exchange-correlation (XC), and kinetic energy terms.
In the WFT-in-DFT embedding approach, the DFT embedding potential is included as an external potential for 
WFT calculations, providing a WFT-level description for one (or more) subsystem while the remaining subsystems and their interactions are seamlessly treated at the DFT level of theory. 

Several groups, including this one, have recently demonstrated that non-additive kinetic energy contributions to the embedding potential can be exactly computed\cite{tfm10jason,tfm11jason,Vis10b,Was11,Car11a} with the use of optimized effective potential (OEP) methods.\cite{Yang03,Yan07,Zhang09,Jacob11,Zha92,Zha93,Zha94} 
In this paper, 
we introduce a simple  technique to improve the robustness of OEP calculations in systems that exhibit small HOMO-LUMO gaps, such as transition metal complexes.  In addition, we derive spin-dependent embedding potentials 
to enable the accurate description of open-shell systems in the WFT-in-DFT embedding framework.
Numerical applications to the van-der-Waals-bound ethylene-propylene dimer
and to the
hexaaquairon(II) transition-metal cation illustrate the applicability of these new techniques and demonstrate the accuracy of the WFT-in-DFT  approach in systems for which conventional density functional theory methods exhibit substantial errors.

\section{Theory}

Like Kohn-Sham (KS)-DFT, DFT embedding provides a formally exact framework for the ground-state electronic structure problem. Here, we review DFT-in-DFT embedding and its basis for WFT-in-DFT calculations.

\subsection{DFT-in-DFT embedding}
  
We begin by considering a closed-shell system in which the total electronic density $\rho_{\textrm{AB}}$ consists of two subsystems,  $\rho_{\textrm{AB}} = \rho_{\textrm{A}} + \rho_{\textrm{B}}$.  The corresponding one-electron orbitals for $\rho_{\textrm{A}}$ and $\rho_{\textrm{B}}$ obey the Kohn-Sham Equations with Constrained Electron Density (KSCED),\cite{Wes06} 
\begin{eqnarray}
\left[-\frac12\nabla^2+v_\textrm{eff}[\rho_\textrm{A},\rho_\textrm{AB};{\bf r}]\right]\phi_i^\textrm{A}({\bf r})=\epsilon_i^\textrm{A}\phi_i^\textrm{A}({\bf r}),  \label{kscedA}\\
\left[-\frac12\nabla^2+v_\textrm{eff}[\rho_\textrm{B},\rho_\textrm{AB};{\bf r}]\right]\phi_j^\textrm{B}({\bf r})=\epsilon_j^\textrm{B}\phi_j^\textrm{B}({\bf r}), \label{kscedB}
\end{eqnarray}
where $i=1,\ldots,N^\textrm{A}$, $j=1,\ldots,N^\textrm{B}$, and 
$N^\textrm{A}$ and $N^\textrm{B}$ are the number of electrons in the respective subsystems. $v_\textrm{eff}$ is the  effective potential for the coupled one-electron equations,
\begin{eqnarray}
v_\textrm{eff}[\rho_\textrm{A},\rho_\textrm{AB};{\bf r}]&=& v^\textrm{KS}_\textrm{eff}[\rho_\textrm{A};{\bf r}] + v_\textrm{emb} (\mathbf{r}),
\end{eqnarray}
where $v^\textrm{KS}_\textrm{eff}[\rho_\textrm{A};{\bf r}]$ is the standard KS potential for subsystem A, and 
\begin{eqnarray}
v_\textrm{emb} (\mathbf{r}) &=&v^\textrm{B}_{\textrm{ne}}(\mathbf{r}) + v_{\textrm{J}}[\rho_\textrm{B}; \mathbf{r}] + v_{\textrm{xc}}[\rho_\textrm{AB}; \mathbf{r}]-  \nonumber\\
&& v_{\textrm{xc}}[\rho_\textrm{A}; \mathbf{r}]  +v_{\textrm{nad}}[\rho_{\textrm{A}},\rho_{\textrm{AB}}; \mathbf{r}].
 \label{emb}
\end{eqnarray}
Here, $v^\textrm{B}_{\textrm{ne}}(\mathbf{r})$ is the nuclear-electron Coulomb potential from the nuclei contained in subsystem B, $v_{\textrm{J}}$ is the Hartree potential, and $v_{\textrm{xc}}$ is the XC potential. In addition to these familiar terms from conventional KS-DFT calculations, DFT embedding introduces the non-additive kinetic potential (NAKP) which properly enforces Pauli exclusion between the subsystem densities.  It is obtained from
 \begin{equation}
 v_{\textrm{nad}}[\rho_{\textrm{A}},\rho_{\textrm{AB}}; \mathbf{r}]  = \left[ \frac{\delta{T_{\textrm{s}}^{\textrm{nad}}{[\rho_{\textrm{A}},\rho_{\textrm{B}}]}}} {\delta{\rho_{\textrm{A}}}} \right] \!\!(\mathbf{r}),
 \label{nakp}
 \end{equation}
 where  $T_{\textrm{s}}^\textrm{nad}[\rho_\textrm{A},\rho_\textrm{B}]\equiv T_{\textrm{s}}[\rho_\textrm{AB}]-T_{\textrm{s}}[\rho_\textrm{A}]-T_{\textrm{s}}[\rho_\textrm{B}]$, and $T_{\textrm{s}}[\rho]$ is the non-interacting kinetic energy functional.  The total energy functional for the full system is then
\begin{eqnarray}
E[\rho_{\textrm{AB}}] &=& T_{\textrm{s}}[\rho_{\textrm{A}}] + T_{\textrm{s}}[\rho_{\textrm{B}}] + T_{\textrm{s}}^{\textrm{nad}}[\rho_{\textrm{A}},\rho_{\textrm{B}}]   + \nonumber\\
&& V_{\textrm{ne}}[\rho_{\textrm{AB}}]  +  J[\rho_{\textrm{AB}}] + E_{\textrm{xc}}[\rho_{\textrm{AB}}],
\label{tenergy}
\end{eqnarray}
where the last three terms on the right-hand side (RHS) are the nuclear-electron Coulomb energy, the Hartree energy, and the XC energy computed over the total density.   
Enforcing $v_\textrm{emb} (\mathbf{r})$ to be identical for all subsystems (see Sec.~III B) leads to a unique partitioning in the DFT embedding formulation, such that the specification of the nuclei and the integer number of electrons in subsystem A and B fully specify the density partitioning.\cite{Was11, Car11a} 
Eqs. 1-\ref{tenergy} are easily generalized to the description of multiple embedded subsystems.  

We have previously demonstrated that by using OEP methods to calculate the NAKP, DFT-in-DFT embedding can accurately describe both weakly and strongly interacting subsystems, including subsystems connected by covalent bonds;\cite{tfm10jason,tfm11jason} and we have shown that this method is computationally feasible for large systems by way of localized approximations to the NAKP.\cite{tfm11jason} More recently, we have introduced a projection approach that completely avoids the NAKP calculation in exact DFT embedding,\cite{tfm12jason} which appears worthy of further investigation.  The OEP-based approach employed here is
appealing because it provides a local embedding potential that is a functional of only the subsystem electronic densities.   

In practice, the KSCED equations (Eq.~\ref{kscedA} and Eq.~\ref{kscedB}) are solved by simply modifying the core Hamiltonian in the self-consistent field (SCF) calculation to include the additional embedding terms. The embedding potential (Eq.~\ref{emb}) can be written in the atomic orbital (AO) basis as 
\begin{eqnarray}
\mathbf{v}_\textrm{emb} &=& \mathbf{v}^\textrm{B}_{\textrm{ne}} + \mathbf{J}[\gamma_\textrm{B}] + \mathbf{v}_{\textrm{xc}}[\gamma_\textrm{AB}]-  \nonumber\\
&& \mathbf{v}_{\textrm{xc}}[\gamma_\textrm{A}]  +\mathbf{v}_{\textrm{nad}}[\gamma_{\textrm{A}},\gamma_{\textrm{AB}}],
 \label{matrixemb}
 \end{eqnarray}
where the various terms on the RHS of this expression correspond to those in Eq.~\ref{emb}.  The subsystem and total AO density matrices in Eq.~\ref{matrixemb} satisfy $\gamma_\textrm{A} + \gamma_\textrm{B} = \gamma_\textrm{AB}$.  It follows that the Fock matrix for subsystem A can be expressed as
\begin{eqnarray}
\mathbf{f}^{\textrm{A in B}} &=& \mathbf{h}^{\textrm{A in B}} +  \mathbf{J}[\gamma_\textrm{A}] + \mathbf{v}_{\textrm{xc}}[\gamma_\textrm{A}],
 \label{matrixfock}
\end{eqnarray}
where 
\begin{eqnarray}
\mathbf{h}^{\textrm{A in B}}= \mathbf{h}^{\textrm{A}} +  \mathbf{v}_\textrm{emb},
 \label{coreainb}
\end{eqnarray}
and $\mathbf{h}^{\textrm{A}}$ is the core Hamiltonian for subsystem A (the kinetic energy plus external potential due to the nuclei in subsystem A).   The Fock matrix  for subsystem B, $\mathbf{f}^{\textrm{B in A}}$, is analogously defined.

 \subsection{WFT-in-DFT embedding}

The embedding potential in Eq.~\ref{emb} describes the subsystem interactions in terms of their corresponding electronic densities. However, the subsystem densities can be computed with any level of theory, thus allowing for the description of one subsystem at the 
(single- or multi-reference) WFT level, while the remaining environment is treated at the DFT level.\cite{carterembedding,tfm12jason,Vis08b,Car11a,Car09,Car01,Wes08,Car06,Car09b}
 Closed-shell WFT-in-DFT embedding simply involves performing a WFT calculation on a given subsystem using the modified core Hamiltonian, $\mathbf{h}^{\textrm{A in B}}$ in Eq.~\ref{coreainb}, that contains the embedding terms due to the environment of the other subsystem.  The WFT-in-DFT energy is then obtained by modifying the DFT energy with respect to subsystem contributions at the WFT level,\cite{Car11a}
\begin{eqnarray}
E_{\textrm{tot}}\left[ \rho^{\textrm{WFT}}_{\textrm{A}}, \rho^{\textrm{DFT}}_{\textrm{B}}  \right] &=& E_{\textrm{AB}}^{\textrm{DFT}}[\rho^{\textrm{DFT}}_{\textrm{AB}}]   \nonumber \\ 
&& \hspace{-25mm} - \left( E_{\textrm{A}}^{\textrm{DFT}}[\rho^{\textrm{DFT}}_{\textrm{A}}] + \int v_\textrm{emb} (\mathbf{r})  \rho^{\textrm{DFT}}_{\textrm{A}} (\mathbf{r})  d{\bf r} \right) \nonumber \\ 
&& \hspace{-25mm} + \left( E_{\textrm{A}}^{\textrm{WFT}}[\rho^{\textrm{WFT}}_{\textrm{A}}] + \int v_\textrm{emb} (\mathbf{r})  \rho^{\textrm{WFT}}_{\textrm{A}} (\mathbf{r})  d{\bf r} \right).  
 \label{eq:energy}
\end{eqnarray}
This expression is easy to evaluate since the terms in the parentheses are just the DFT and WFT energies of subsystem A performed using the modified core Hamiltonian, $\mathbf{h}^{\textrm{A in B}}$.  
Just as DFT-in-DFT embedding is an exact theory for the case of an exact DFT XC functional, WFT-in-DFT embedding is an exact theory for the case of an exact DFT XC functional and a full configuration interaction (FCI) WFT description.\cite{Wes08} 
%


   \section{Methods of Implementation}  

Here, we describe techniques to improve the accuracy and convergence of both DFT-in-DFT and WFT-in-DFT calculations. First, a description of open-shell DFT embedding is developed to incorporate the effects of spin-dependence in the embedding potential.  Then, implementation of the OEP calculation is discussed, and an orbital occupation constraint is introduced to enable robust DFT-in-DFT and WFT-in-DFT embedding calculations for systems with low-lying virtual orbitals, such as transition metal complexes.

  \subsection{Embedding for open-shell systems}
 
 For an open-shell embedded subsystem, the $\alpha$ and $\beta$ electrons generally experience different embedding potentials due to differing non-additive XC and NAKP contributions. Previous WFT-in-DFT implementations for open-shell systems have in practice neglected this difference, effectively assuming that the spin polarization is localized within the WFT subsystem.\cite{Vis08b,Car06}  In this study, we show that effects due to spin-dependent potentials are substantial and easily included via separate $\alpha$ and $\beta$ embedding potentials.   We develop approaches to utilize both restricted and unrestricted open-shell orbital formulations in WFT calculations.

 \subsubsection{Open-shell DFT-in-DFT embedding}

We begin by considering an open-shell system for which the total electronic density is comprised of the $\alpha$ and $\beta$ density 
of the two subsystems,
 $\rho_{\textrm{AB}} = \rho_{\textrm{A}}^{\alpha}+\rho_{\textrm{A}}^{\beta}+\rho_{\textrm{B}}^{\alpha} +\rho_{\textrm{B}}^{\beta} $.  The effective potential for the unrestricted spin orbitals\cite{tfm10jason} is 
\begin{eqnarray}
v^{\alpha}_\textrm{eff}[\rho_{\textrm{A}}^{\alpha},\rho_{\textrm{A}}^{\beta},\rho_{\textrm{B}}^{\alpha},\rho_{\textrm{B}}^{\beta};{\bf r}]&=& v^{\textrm{KS},\alpha}_\textrm{eff}[\rho_{\textrm{A}}^{\alpha},\rho_{\textrm{A}}^{\beta};{\bf r}] + v^{\alpha}_\textrm{emb} (\mathbf{r}),
\end{eqnarray}
where $v^{\textrm{KS},\alpha}_\textrm{eff}[\rho_{\textrm{A}}^{\alpha},\rho_{\textrm{A}}^{\beta};{\bf r}]$ is the standard KS effective potential for the unrestricted (U)KS orbitals, and $v^{\alpha}_\textrm{emb} (\mathbf{r})$ is a spin-dependent embedding potential applied only to the $\alpha$-spin electrons, 
\begin{eqnarray}
v^{\alpha}_\textrm{emb} (\mathbf{r})&=&v^\textrm{B}_{\textrm{ne}}(\mathbf{r}) + v_{\textrm{J}}[\rho_\textrm{B}; \mathbf{r}] + v_{\textrm{xc}}^{\alpha}
[\rho_{\textrm{AB}}^{\alpha},\rho_{\textrm{AB}}^{\beta};{\bf r}] -  \nonumber\\
&& v_{\textrm{xc}}^{\alpha}[\rho_{\textrm{A}}^{\alpha},\rho_{\textrm{A}}^{\beta}; \mathbf{r}]  +v_{\textrm{nad}}^{\alpha}[\rho_{\textrm{A}}^{\alpha},\rho_{\textrm{AB}}^{\alpha}; \mathbf{r}].
\label{uemb}
\end{eqnarray}
The corresponding quantities for the $\beta$-spin electrons are analogously defined.  The total energy functional for the full open-shell system is then
\begin{eqnarray}
E[\rho_{\textrm{AB}}] &=& T_{\textrm{s}}[\rho^{\alpha}_{\textrm{A}}] + T_{\textrm{s}}[\rho^{\alpha}_{\textrm{B}}] + T_{\textrm{s}}^{\textrm{nad}}[\rho^{\alpha}_{\textrm{A}},\rho^{\alpha}_{\textrm{B}}]   +  \nonumber\\ 
&& T_{\textrm{s}}[\rho^{\beta}_{\textrm{A}}] + T_{\textrm{s}}[\rho^{\beta}_{\textrm{B}}] + T_{\textrm{s}}^{\textrm{nad}}[\rho^{\beta}_{\textrm{A}},\rho^{\beta}_{\textrm{B}}]   +  \nonumber\\ 
&& V_{\textrm{ne}}[\rho_{\textrm{AB}}]  +  J[\rho_{\textrm{AB}}] + E_{\textrm{xc}}[\rho_{\textrm{AB}}^{\alpha},\rho_{\textrm{AB}}^{\beta}].
\label{ostenergy}
\end{eqnarray}
\noindent
Separate OEP calculations are performed over the $\alpha$ and $\beta$ spin-densities for the exact calculation of the NAKP, which allows for numerically exact unrestricted open-shell DFT embedding (U-DFT-in-DFT).\cite{tfm10jason}

In practice, we solve for the unrestricted spin orbitals by adding the spin-dependent embedding potentials to the $\alpha$ and $\beta$ Fock matrices.   The $\alpha$ and $\beta$ embedding potential can be written in the AO basis as
\begin{eqnarray}
\mathbf{v}^{\xi}_\textrm{emb} &=&\mathbf{v}^\textrm{B}_{\textrm{ne}} + \mathbf{J}[\gamma_\textrm{B}] + \mathbf{v}_{\textrm{xc}}^{\xi}
[\gamma_{\textrm{AB}}^{\alpha},\gamma_{\textrm{AB}}^{\beta}] -  \nonumber\\
&& \mathbf{v}_{\textrm{xc}}^{\xi}[\gamma_{\textrm{A}}^{\alpha},\gamma_{\textrm{A}}^{\beta}]  +\mathbf{v}_{\textrm{nad}}^{\xi}[\gamma_{\textrm{A}}^{\xi},\gamma_{\textrm{AB}}^{\xi}],    
\label{spinmatrixemb}
\end{eqnarray}
where $\xi \in \{\alpha, \beta\}$, and the corresponding Fock matrices are 
\begin{eqnarray}
\mathbf{f}^{\xi,\textrm{A in B}} &=& \mathbf{h}^{\textrm{A}} +  \mathbf{J}[\gamma_\textrm{A}] + \mathbf{v}_{\textrm{xc}}^{\xi}[\gamma_{\textrm{A}}^{\alpha},\gamma_{\textrm{A}}^{\beta}]  +\mathbf{v}^{\xi}_\textrm{emb}.\label{matrixfockspin}
\end{eqnarray}
\noindent
The unrestricted spin orbitals for subsystem A are then obtained by separately diagonalizing $\mathbf{f}^{\alpha,\textrm{A in B}}$ and $\mathbf{f}^{\beta,\textrm{A in B}}$ in the usual way.

Practical implementations for performing OEP calculations using restricted open-shell orbitals have yet to be developed. We thus introduce a simple, approximate scheme for restricted open-shell DFT embedding (RO-DFT-in-DFT). In this approach, a U-DFT-in-DFT calculation is first performed, and the  embedding potentials $\mathbf{v}^{\alpha}_\textrm{emb}$ and $\mathbf{v}^{\beta}_\textrm{emb}$ are constructed using Eq.~\ref{spinmatrixemb}. Then, $\mathbf{f}^{\alpha,\textrm{A in B}}$ and $\mathbf{f}^{\beta,\textrm{A in B}}$ are constructed using Eq.~\ref{matrixfockspin}, and the usual RO approach is employed to obtain subsystem orbitals that are spatially identical for the $\alpha$ and $\beta$ electrons.
Specifically,  $\mathbf{f}^{\alpha,\textrm{A in B}}$ is diagonalized to obtain a set of occupied $\alpha$ spin orbitals, \{$\phi^{\alpha,\textrm{A}}_{\textrm{occ}}$\}, and $\mathbf{f}^{\beta,\textrm{A in B}}$ is then projected into the space of the occupied $\alpha$ spin orbitals using
 \begin{eqnarray}
\tilde{\mathbf{f}}^{\beta,\textrm{A in B}} = \mathbf{c}^\textrm{T}\mathbf{f}^{\beta,\textrm{A in B}}\mathbf{c},
\end{eqnarray}
where $\mathbf{c}$ is the matrix with columns  comprised of the AO coefficients for  \{$\phi^{\alpha,\textrm{A}}_{\textrm{occ}}$\}.  Finally, the projected Fock matrix, $\tilde{\mathbf{f}}^{\beta,\textrm{A in B}}$, is diagonalized to obtain the set of RO orbitals,  \{$\phi^{\textrm{A}}_{\textrm{occ}}$\}, with the first $N^{\beta,\textrm{A}}$ orbitals doubly occupied and with orbitals $N^{\beta,\textrm{A}}+1,\ldots,N^{\alpha,\textrm{A}}$ singly occupied, where $N^{\alpha,\textrm{A}}$ and $N^{\beta,\textrm{A}}$ indicate the number of $\alpha$ and $\beta$ electrons in subsystem A.  Although the second and third terms on the RHS of Eq.~\ref{matrixfockspin} are updated at each iteration of the RO-DFT-in-DFT calculation, we leave the embedding potentials unchanged to avoid performing OEP calculations using restricted open-shell orbitals. The RO-DFT-in-DFT energy for the total density is calculated using Eq.~\ref{ostenergy}.

Several different schemes have been proposed to calculate the embedding potential for open-shell subsystems while neglecting its spin-dependence.\cite{Vis08b,Car11a,Car06}  
These approaches generally assume that interactions between the subsystems can be described by a single embedding potential.  For example, in systems with an even number of electrons, the embedding potential, $\mathbf{v}_\textrm{emb}$ in Eq.~\ref{matrixemb}, can be obtained assuming that each embedded subsystem is closed-shell, 
and then the open-shell subsystem is calculated 
using $\mathbf{v}^{\alpha}_\textrm{emb} = \mathbf{v}^{\beta}_\textrm{emb} = \mathbf{v}_\textrm{emb}$. 
In this approach, Eq.~\ref{matrixfockspin} is solved self-consistently while $\mathbf{v}^{\alpha}_\textrm{emb}$ and $\mathbf{v}^{\beta}_\textrm{emb}$ are held fixed, and the final DFT-in-DFT energy is calculated using Eq.~\ref{ostenergy}.  
This spin-independent description of the embedding potential can be used  with either an unrestricted treatment of the open-shell subsystem (U-DFT-in-DFT-CS) or a restricted treatment of the open-shell subsystem (RO-DFT-in-DFT-CS); we later employ the approach to compare with the previously described methods (U-DFT-in-DFT and RO-DFT-in-DFT) that include spin-dependence in the embedding potential.


 \subsubsection{Open-shell WFT-in-DFT embedding}
 
Unrestricted open-shell WFT-in-DFT (U-WFT-in-DFT) calculations are performed by first computing unrestricted Hartree-Fock (UHF) orbitals in the spin-dependent embedding potential, and then using these orbitals for a post-HF WFT calculation.  The $\alpha$ and $\beta$ Fock matrices for the calculation of the UHF orbitals are
\begin{eqnarray}
\mathbf{f}^{\xi,\textrm{A in B}} &=& \mathbf{h}^{\textrm{A}} +  \mathbf{J}[\gamma_\textrm{A}] + \mathbf{K}^{\xi}[\gamma_{\textrm{A}}^{\xi}]  + \mathbf{v}^{\xi}_\textrm{emb} ,   \label{matrixfockspinHF} 
\end{eqnarray}
where $\xi \in \{\alpha, \beta\}$, $\mathbf{K}$ is the HF exchange matrix, and the embedding potentials (Eq.~\ref{spinmatrixemb}) are obtained from a U-DFT-in-DFT embedding calculation. 
The total energy is evaluated using
\begin{eqnarray}
 \hspace{-5mm} E_{\textrm{tot}}\left[ \rho^{\textrm{WFT}}_{\textrm{A}}, \rho^{\textrm{DFT}}_{\textrm{B}}  \right] &=& E_{\textrm{AB}}^{\textrm{DFT}}[\rho^{\textrm{DFT}}_{\textrm{AB}}]   \nonumber \\ 
&& \hspace{-35mm} -\left( E_{\textrm{A}}^{\textrm{DFT}}[\rho^{\textrm{DFT}}_{\textrm{A}}] +  \sum_{\xi \in \{\alpha,\beta \} } \int v^{\xi}_\textrm{emb} (\mathbf{r})  \rho^{\xi,\textrm{DFT}}_{\textrm{A}} (\mathbf{r})  d{\bf r} \right) \nonumber \\ 
&& \hspace{-35mm} + \left( E_{\textrm{A}}^{\textrm{WFT}}[\rho^{\textrm{WFT}}_{\textrm{A}}] + \sum_{\xi \in \{\alpha,\beta \} }  \int v^{\xi}_\textrm{emb} (\mathbf{r})  \rho^{\xi,\textrm{WFT}}_{\textrm{A}} (\mathbf{r})  d{\bf r} \right).
\label{osenergy}
\end{eqnarray}

Restricted open-shell WFT-in-DFT (RO-WFT-in-DFT) calculations are performed by solving for restricted open-shell HF (ROHF) orbitals in the spin-dependent embedding potential. Just as in RO-DFT-in-DFT embedding, a U-DFT-in-DFT calculation is first performed, and the embedding potentials $\mathbf{v}^{\alpha}_\textrm{emb}$ and $\mathbf{v}^{\beta}_\textrm{emb}$ are constructed using  Eq.~\ref{spinmatrixemb}.  Then, the Fock matrices in Eq.~\ref{matrixfockspinHF} are constructed and the usual approach is employed to obtain RO orbitals; the second and third terms on the RHS of Eq.~\ref{matrixfockspinHF} are updated at each iteration while the embedding potential is left unchanged.  The ROHF orbitals 
are used in the post-HF WFT calculation, and the total energy is then evaluated using Eq.~\ref{osenergy}.
We note that for the RO-WFT-in-DFT energy calculation, the term $E_{\textrm{A}}^{\textrm{DFT}}[\rho^{\textrm{DFT}}_{\textrm{A}}]$, is evaluated using the ROKS-DFT energy.

   \subsection{Optimized effective potential}  

As seen in Eqs.~\ref{nakp} and \ref{tenergy}, DFT embedding requires computation of both the kinetic energy, $T_{\textrm{s}}[\rho_\textrm{AB}]$, and its functional derivative.  However, since the explicit functional form for the kinetic energy is unknown, OEP methods are needed to obtain these terms exactly.  

The OEP is the local potential for which solution of the one-electron equations
 \begin{eqnarray}
 \left[ -\frac12\nabla^2 + v_\textrm{OEP}({\bf r}) \right]  \phi_i &=&  \epsilon_i \phi_i
 \label{eig}
\end{eqnarray}
yields orbitals that correspond to a given target density while minimizing the non-interacting kinetic energy. 
  A variety of methods for determining such potentials from an input target density have been developed.\cite{Yang03,Yan07,Zhang09,Jacob11,Zha92,Zha93,Zha94} Calculations reported here employ the direct optimization procedure developed by Wu and Yang,\cite{Yang03,Yan07} in which the kinetic energy is obtained via the unconstrained maximization
 \begin{equation}
T_{\textrm{s}}[\rho_\textrm{in}] = \max_{ v_\textrm{OEP}({\bf r})} \left\{ W_{\textrm{s}} \left[ \Psi_{\textrm{det}}, v_\textrm{OEP}({\bf r}) \right] \right\},
 \end{equation}
where
 \begin{eqnarray}
W_{\textrm{s}} \left[ \Psi_{\textrm{det}}, v_\textrm{OEP}({\bf r}) \right] &=& 2 \sum_i^{\frac{N}{2}}\bra{\phi_i} -\frac12\nabla^2 \ket{\phi_i}  \nonumber\\
&& +  \int \left( \rho_\textrm{OEP}({\bf r}) - \rho_\textrm{in}({\bf r}) \right) v_\textrm{OEP}({\bf r}) d{\bf r} \nonumber\\
&& -~\zeta || \nabla v_\lambda({\bf r}) ||^2,
\label{ws}
\end{eqnarray}
\noindent
and
 \begin{eqnarray}
v_\textrm{OEP}({\bf r}) &=& v^\textrm{KS}_\textrm{eff}[ \rho_\textrm{in};{\bf r}]  + v_\lambda({\bf r}). \label{voep} 
\end{eqnarray}
\noindent
In these equations, $v_\lambda ({\bf r}) = \sum_t b_t g_t({\bf r})$, 
\{$g_t({\bf r})$\} comprise an auxiliary basis set for the potential, $b_t$ are the corresponding expansion coefficients, and $\zeta$ is a regularization parameter.\cite{Yan07} Maximization of $W_{\textrm{s}}$ utilizes the Newton method for optimization with a back-tracking line search in the expansion coefficients,\cite{Mordecai_03} 
 \begin{equation}
 \mathbf{b}^{(i+1)} =  \mathbf{b}^{(i)} + \tau \mathbf{H}^{-1} \mathbf{g},
 \label{backtrack}
\end{equation}
where $i$ is the iteration number, ${\bf H}$ and ${\bf g}$ are the Hessian and gradient of $W_s$, respectively, and $\tau \in$ [0,1] is the step-size in the line search.  

In practice, to obtain the embedding potential, 
we do not explicitly calculate the NAKP for each subsystem.  
Instead, for closed shell subsystems, we directly update the  embedding potential (Eq.~\ref{emb}) at each iteration of the KSCED equations using\cite{Was11}
 \begin{equation}
 v^{(i+1)}_\textrm{emb}({\bf r}) = v^{(i)}_\textrm{emb}({\bf r}) - \theta v_\lambda (\mathbf{r}),
 \label{damp}
\end{equation}
 where $\theta \in$ [0,1] is a damping coefficient. 
By construction, 
 the embedding potential for each subsystem is 
 identical at every iteration. 
 The KSCED equations are initialized using $v^{(0)}_\textrm{emb}({\bf r})=0$,  such that the initial guess for the NAKP for subsystem A exactly cancels the remaining terms in Eq.~\ref{emb} (and the initial guess for the NAKP for subsystem B likewise cancels the corresponding terms). 
 Upon convergence of the KSCED equations, 
 $v_\lambda (\mathbf{r})=0$, 
 $v^\textrm{KS}_\textrm{eff}[ \rho_\textrm{AB};{\bf r}] = v^\textrm{KS}_\textrm{eff}[ \rho_\textrm{KS};{\bf r}]$, 
 and 
 $v_\textrm{emb}({\bf r})=v^{(i)}_\textrm{emb}({\bf r})$. 
Enforcing the embedding potential 
to be identical for all subsystems  leads to a unique partitioning of the subsystem densities.\cite{Was11, Car11a}

For open-shell calculations, we similarly update the spin-dependent embedding potential (Eq.~\ref{uemb}); 
the OEP obtained for a given spin density is
\begin{equation}
 v^{\xi}_\textrm{OEP} (\mathbf{r}) = v^{\textrm{KS},\xi}_\textrm{eff}[(\rho_{\textrm{A}}^{\alpha}+\rho_{\textrm{B}}^{\alpha}),(\rho_{\textrm{A}}^{\beta}+\rho_{\textrm{B}}^{\beta});{\bf r}] + v^{\xi}_\lambda (\mathbf{r}),
\end{equation}
and as in Eq.~\ref{damp}, $v^{\xi}_\lambda (\mathbf{r})$ is used to update $v^{\xi}_\textrm{emb}({\bf r})$ at each iteration.

Finally, we note that XC functionals that include a fraction of the exact exchange can be employed in DFT embedding via the OEP calculation.  
The HF exchange matrix, $\mathbf{K}$, is evaluated using $\gamma_\textrm{OEP}$, the OEP density matrix in the AO basis.
For DFT-in-DFT embedding, the exchange energy is calculated using
 \begin{eqnarray}
E_\textrm{X} \left[\gamma_\textrm{OEP} , \gamma_\textrm{in} \right] &=& -\mathrm{tr} \left(  \gamma_\textrm{OEP} \mathbf{K}[\gamma_\textrm{OEP}] \right) \nonumber \\
&& + \mathrm{tr} \left(  (\gamma_\textrm{OEP} - \gamma_\textrm{in}) \mathbf{K}[\gamma_\textrm{OEP}] \right),
\label{exx}
\end{eqnarray}
where the second term on the RHS corrects the exchange energy for small numerical differences between $\gamma_\textrm{OEP}$ and  $\gamma_\textrm{in}$. For calculations on the low-spin state of the hexaaquairon(II) cation, this correction is found to be as large as  20 kcal/mol; however, the correction is not required for the evaluation of the WFT-in-DFT energy (Eq.~\ref{eq:energy}), since $E_{\textrm{AB}}^{\textrm{DFT}}[\rho^{\textrm{DFT}}_{\textrm{AB}}]$ is obtained directly from a KS-DFT calculation on the full system.

   \subsection{Orbital-occupation freezing}  

For $W_\textrm{s}$ to be a concave function of $v^\textrm{OEP}({\bf r})$, it is necessary\cite{Yang03} that the orbitals used to construct $\rho_{\textrm{OEP}}$ in Eq.~\ref{ws} correspond to the lowest eigenvalues of Eq.~\ref{eig}.
However, this can be problematic for systems with small energy differences between the occupied and virtual orbitals, where small changes in $v_\textrm{OEP}({\bf r})$ can alter the relative ordering of the orbitals.

\begin{figure}
  \begin{center}
    \includegraphics[angle=0,width=8.5cm]{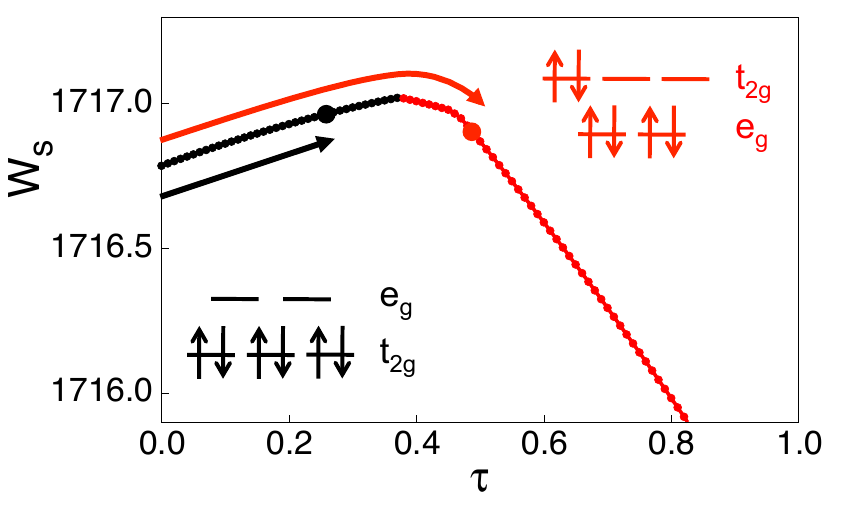}
  \end{center}
  \vspace{-20pt}
\caption{ 
An illustrative Newton step in the OEP calculation for  the low-spin hexaaquairon(II) cation, performed with (black) and without (red) the orbital-occupation-freezing technique.
The technique ensures that  correct  orbitals remain occupied throughout the maximization of $W_\textrm{s}$.  See text for details.
 }
\label{fig:oep}
\end{figure}

To illustrate this issue, Fig.~\ref{fig:oep} shows the line search for an illustrative Newton step in an OEP calculation for the low-spin hexaaquairon(II) cation.
$W_\textrm{s}$ is plotted as a function of $\tau$, where $\tau$ is the step-size in Eq.~\ref{backtrack}.  
For any step-size larger than $\tau = 0.38$ in this case, the orbital occupancy changes from one in which only t$_{\textrm{2g}}$-like $d$ orbitals are occupied to one in which e$_{\textrm{g}}$-like $d$ orbitals are occupied.   In traditional back-tracking line searches, any step which increases $W_\textrm{s}$ would be accepted, including the $\tau = 0.5$ step indicated with the red arrow. However, this step is problematic since the Hessian and gradient of $W_\textrm{s}$ for the next Newton step would be evaluated using a density that corresponds to the wrong orbitals.  The net results are poor convergence 
and incorrect solutions for the OEP.

We introduce a simple method to alleviate this problem by modifying the back-tracking line search.  
Reference ($\tau$=0) orbitals are computed from Eq.~\ref{eig} using $v_\textrm{OEP}({\bf r}) = v^\textrm{KS}_\textrm{eff}[ \rho_\textrm{AB};{\bf r}]$, and for any proposed step-size $\tau$, the corresponding orbitals are computed using $v_\textrm{OEP}({\bf r}) = v^\textrm{KS}_\textrm{eff}[ \rho_\textrm{AB};{\bf r}] + v_\lambda ({\bf r})$.  The proposed step is rejected if the overlap between these two sets of orbitals is less than 0.5, regardless of the change in $W_\textrm{s}$; otherwise, it is subjected to the usual criteria of the back-tracking line search algorithm.   Upon rejection, the step-size $\tau$ is reduced by a factor of 2. 
This technique ensures that the correct orbitals remain occupied throughout the maximization of $W_\textrm{s}$.
In Fig.~\ref{fig:oep}, 
the proposed step indicated by the red arrow is rejected, whereas the proposed shorter step indicated by the black arrow is accepted; not only is the value of $W_\textrm{s}$ increased, but the correct orbitals remain occupied.  By utilizing this technique, we found that the maximization of $W_\textrm{s}$ typically requires less than 20 Newton steps for the low spin state of the hexaaquairon(II) cation, whereas the optimization failed to converge without the use of orbital-occupation freezing.

\subsection{Computational Details}

The DFT embedding methods employed here are all implemented in the development version of the Molpro software package.\cite{Molpro}  All calculations employ the supermolecular basis set convention, in which the molecular orbitals for each subsystem are described in the AO basis  for the full system.\cite{Wes97}  Calculations on the ethylene-propylene dimer use the aug-cc-pVTZ orbital basis set for the carbon atoms and the aug-cc-pVDZ orbital basis set for the hydrogen atoms. Calculations on the hexaaquairon(II) cation use the aug-cc-pVTZ orbital basis set for the iron atom and the aug-cc-pVDZ orbital basis set for the hydrogen and oxygen atoms. For the auxiliary basis set used in the OEP calculations, we employ atom-centered Gaussian basis functions ($g_t({\bf r}) = N_t e^{-\lambda_t \mathbf{r}^2}$, where $N_t$ is the normalization constant) for which the coefficient $\lambda_t$ assumes values of $2^{n}$, where $n=n_{\textrm{min}},n_{\textrm{min}}+2,\dots,n_{\textrm{max}}-2,n_{\textrm{max}}$. Calculations on the ethylene-propylene dimer  employ the basis set for which the $s$-type functions for the carbon and hydrogen atoms span \{$n_{\textrm{min}},n_{\textrm{max}}$\} = \{-4, 4\}, and the $p$-type functions for the carbon and hydrogen atoms span \{-2, 2\}. 
Calculations for the hexaaquairon(II) cation employ the basis set for which the $s$-type functions for the iron atom span \{-4, 6\}, the $p$-type functions for the iron atom span \{-4, 6\}, the $d$-type functions for the iron atom span \{-2, 2\}, the $s$-type functions for the oxygen atoms span \{-4, 6\}, the $p$-type functions for the oxygen atoms span \{-2, 4\}, the $s$-type functions for the hydrogen atoms span  \{-4, 4\}, and the $p$-type functions for the hydrogen atoms span \{-2, 2\}.  
For all systems, the finite auxiliary basis set for the OEP calculations was confirmed to introduce a difference of less than 1 kcal/mol between the total energy computed using KS-DFT and either closed-shell or unrestricted open-shell DFT-in-DFT embedding.
%
The regularization parameter used in the OEP calculations is set to $\zeta = 10^{-3}$; smaller values were tested on the ethylene-propylene dimer and the hexaaquairon(II) cation and were found to have only a small ($\mathcal{O} (\mu\mathrm{Hartree}))$ effect on the total DFT-in-DFT energy.  

The KSCED equations are initialized with subsystem densities comprised of the superposition of HF atomic densities and with 
 $v_\textrm{emb}({\bf r})=0$; 
different initial guesses for the embedding potential were tested on the hexaaquairon(II) cation and were found to yield  similar final embedding potentials with only  small ($\mathcal{O} (10~ \mu\mathrm{Hartree})$) changes in the total DFT-in-DFT energy.

\section{Results}

\subsection{The Ethylene-Propylene Dimer: WFT-in-DFT Embedding}

\begin{figure} 
  \begin{center}
    \includegraphics[angle=0,width=8.5cm]{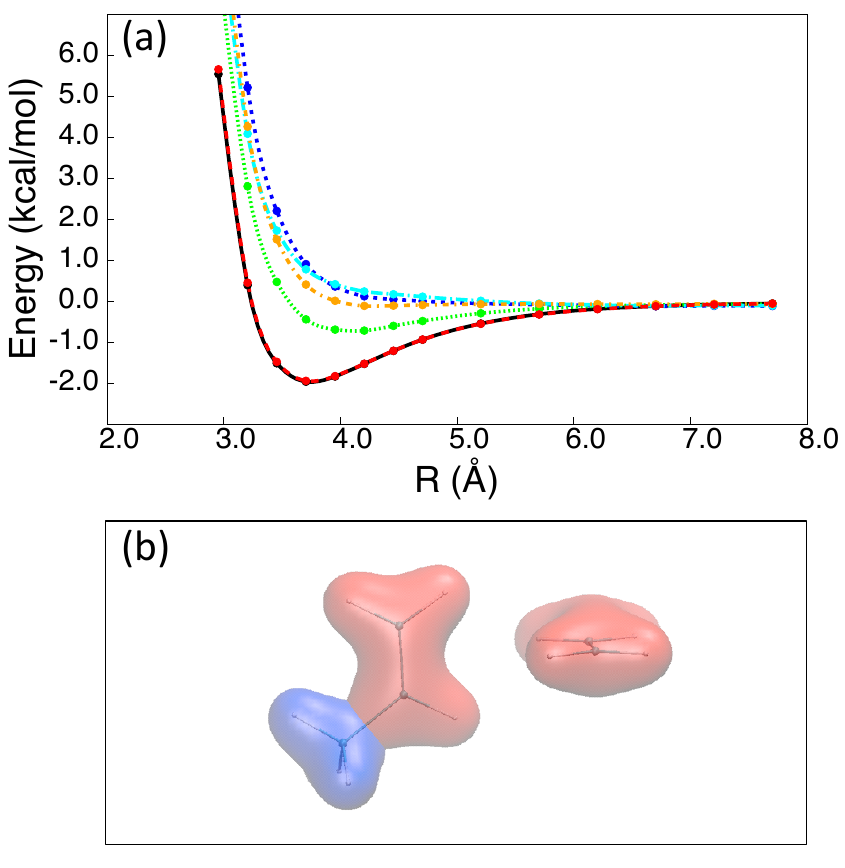}
  \end{center}
  \vspace{-20pt}
\caption{ WFT-in-DFT embedding for the ethylene-propylene dimer. (a) The ethylene-propylene dissociation curve, obtained using CCSD(T)-in-B3LYP (red) and KS-DFT with PBE (green), B3LYP (orange), B-LYP (blue) and B88-P86 (cyan) for the XC functional.  Also included are the reference CCSD(T) results (black), which are graphically indistinguishable from the CCSD(T)-in-B3LYP results.  The curves are vertically shifted to align at infinite separation. 
(b) Isosurface plots indicate the subsystem partitioning for the ethene-propene dimer calculations.  
The red isosurface indicates the density of the 32 electrons associated with the C$_2$H$_4$-C$_2$H$_3$- moiety, and the blue isosurface indicates the density of  the 8 electrons associated with the -CH$_3$ moiety. The isosurface plot corresponds to an electronic density of 0.05 a.u.
}
\label{fig:pi_plot}
\end{figure}

The ethylene-propylene dimer is a prototypical system for which quantum embedding methods, such as QM/MM or ONIOM, may be employed. It exhibits a weak $\pi-\pi$ interaction that is difficult to address with conventional KS-DFT methods, while also exhibiting a spectator -CH$_3$ moiety that contributes little to the interaction energy while substantially increasing the cost of the high-level calculation. However, unlike the QM/MM treatment of subsystems, the interactions between the $\pi-\pi$ system and the -CH$_3$ moiety can be treated seamlessly using WFT-in-DFT embedding, as is now demonstrated. 

Fig.~\ref{fig:pi_plot}(a) presents the ethylene-propylene dimer dissociation curve plotted as a function of the distance between the ethylene and propylene $\pi$ bonds, with the equilibrium dimer geometry obtained via minimization at the MP2 level of theory.  Other geometries along the curve are obtained by displacing the two molecules along the vector formed between the midpoints of the two C=C bonds, while fixing all other internal coordinates.  The relative energies are plotted by aligning each curve at infinite separation.  The full CCSD(T) calculation (black) shows a binding energy of 2.0 kcal/mol.  KS-DFT calculations using the PBE\cite{Ernzerhof96,Ernzerhof97} (green), B3LYP\cite{Becke1993} (orange), B-LYP\cite{Becke1988,Par97} (blue) and B88-P86\cite{Becke1988,Perdew1986} (cyan) XC functionals illustrate the difficulty in describing dispersion interactions using KS-DFT.  The PBE functional underestimates the binding energy by 1.3 kcal/mol, while the rest of the XC functionals fail to capture any of the attractive interactions.

Finally, the red curve in Fig.~\ref{fig:pi_plot}(a) presents the results of 
 WFT-in-DFT embedding, using a subsystem partitioning in which the 32 electrons associated with the $\pi$ system (C$_2$H$_4$-C$_2$H$_3$-, red in Fig.~\ref{fig:pi_plot}(b)) are treated at the WFT level of theory and the remaining 8 electrons in the -CH$_3$ moiety are treated at the DFT level of theory.  We employ CCSD(T) for the WFT and the B3LYP XC functional for the DFT (i.e., CCSD(T)-in-B3LYP).  Fig.~\ref{fig:pi_plot}(a) shows excellent agreement between the CCSD(T) (black) and CCSD(T)-in-B3LYP (red) calculations;  these curves, which are graphically indistinguishable, differ by less than 0.10 kcal/mol through the entire range of distances.  We have confirmed that this level of accuracy is maintained with different XC functionals used for the DFT;   
  specifically, CCSD(T)-in-(B-LYP) energies differ from the CCSD(T) results by less than 0.20 kcal/mol throughout the entire curve. 
 These results illustrate that WFT-in-DFT embedding can be used to systematically improve DFT results and to avoid embedding errors while partitioning across covalent bonds.

\subsection{The hexaaquairon(II) cation}

We now present DFT-in-DFT and WFT-in-DFT calculations for the high-spin [$^5\textrm{T}_{2\textrm{g}}:(\textrm{t}_{2\textrm{g}})^4(\textrm{e}_\textrm{g})^2$] and low-spin [$^1\textrm{A}_{1\textrm{g}}:(\textrm{t}_{2\textrm{g}})^6(\textrm{e}_\textrm{g})^0$] states of the hexaaquairon(II) cation, a system that presents challenges due to the presence of low-lying unoccupied orbitals, the important role of unpaired electrons, and the relatively large number of electrons (84 e$^-$) in the full system.
First, we test the accuracy of DFT-in-DFT embedding for the various treatments of the open-shell embedding potential described earlier. 
We then employ WFT-in-DFT calculations to investigate the low-spin/high-spin energy splitting and the ligation energy for this transition metal complex.

\subsubsection{DFT-in-DFT embedding}

\begin{figure}[h!]  
  \begin{center}
    \includegraphics[angle=0,width=8.5cm]{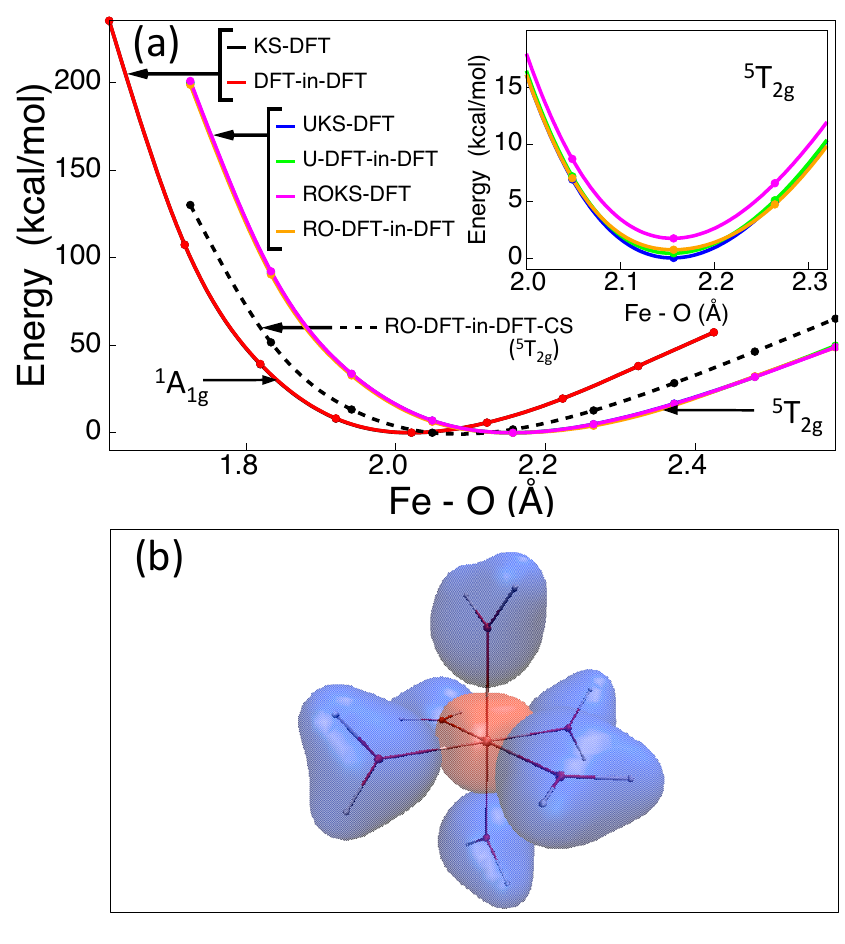}
  \end{center}
  \vspace{-20pt}
\caption{ DFT-in-DFT embedding for the hexaaquairon(II) cation.  
(a) The potential energy curve for the simultaneous dissociation of the six H$_2$O ligands.  
All curves in the main panel are vertically shifted to share a common minimum energy; they are not horizontally shifted.  
The dissociation curves for the low-spin ($^1\textrm{A}_{1\textrm{g}}$) state  
 obtained using KS-DFT (black) and DFT-in-DFT (red)   are graphically indistinguishable.  
 The dissociation curves or the high-spin ($^5\textrm{T}_{2\textrm{g}}$) state obtained using UKS-DFT (blue), U-DFT-in-DFT (green),  ROKS-DFT (magenta), and RO-DFT-in-DFT (orange) are likewise graphically indistinguishable.
 The inset  shows these four high-spin potential energy curves, with each curve vertically shifted only by the UKS-DFT minimum energy of $-$1721.693423 Hartree.
 The dashed black dissociation curve in the main panel is obtained using the RO-DFT-in-DFT-CS method, which neglects spin-dependence in the embedding potential. 
 (b) 
  Isosurface plots indicate the subsystem partitioning for the hexaaquairon(II) cation.  
The red isosurface indicates the density of the 24 electrons associated with the Fe atom, and the blue isosurface indicates the density of  the 60 electrons associated with the six  H$_2$O ligands. The isosurface plot corresponds to an electronic density of 0.05 a.u.
 }
\label{fig:Fe_fig}
\end{figure}

Fig.~\ref{fig:Fe_fig}(a) presents the potential energy curve for the simultaneous dissociation of all six H$_2$O ligands of the hexaaquairon(II) cation, plotted as a function of the average iron-oxygen distance. The equilibrium geometries for the low-spin  [$^1\textrm{A}_{1\textrm{g}}:(\textrm{t}_{2\textrm{g}})^6(\textrm{e}_\textrm{g})^0$] and high-spin [$^5\textrm{T}_{2\textrm{g}}:(\textrm{t}_{2\textrm{g}})^4(\textrm{e}_\textrm{g})^2$] states are obtained using KS-DFT energy minimization with the B3LYP XC functional; 
all other geometries are obtained by uniformly stretching the iron-oxygen distances in the complex, keeping all other internal coordinates unchanged.  
All KS-DFT and DFT-in-DFT embedding results reported in this section are obtained using the B3LYP XC functional.
The curves in the main panel of Fig.~\ref{fig:Fe_fig}(a) are vertically shifted to share a common minimum value; they are not horizontally shifted.  The high-spin state is lower in energy and exhibits a longer average iron-oxygen distance than the low-spin state.

We perform DFT-in-DFT embedding using a subsystem partitioning in which the 24 electrons associated with the iron center comprise one subsystem (red in Fig.~\ref{fig:Fe_fig}(b)) and the remaining 60 electrons associated with the six water ligands comprise a second subsystem (blue in Fig.~\ref{fig:Fe_fig}(b)).  
For the low-spin state, Fig.~\ref{fig:Fe_fig}(a) demonstrates good numerical agreement between DFT-in-DFT (red) and KS-DFT (black); the relative energies differ by less than 0.6 kcal/mol throughout the range of reported internuclear distances.

For the high-spin state of the hexaaquairon(II) cation, Fig.~\ref{fig:Fe_fig}(a) shows that the UKS-DFT and ROKS-DFT methods are in good agreement with each other, as well as with the corresponding U-DFT-in-DFT and RO-DFT-in-DFT embedding approaches described in Sec. III A 1. The U-DFT-in-DFT calculation accurately reproduces the relative energies obtained from UKS-DFT to within 0.4 kcal/mol throughout the attractive branch of the curve and to within 0.8 kcal/mol at shorter distances. The RO-DFT-in-DFT calculation reproduces the relative energy obtained from ROKS-DFT to within 1.0 kcal/mol throughout the attractive branch of the curve and to within 2.2 kcal/mol at shorter distances. 

The inset of Fig.~\ref{fig:Fe_fig}(a) shows the various potential energy curves computed for the high-spin state of the hexaaquairon(II) cation, with each curve vertically shifted by only the UKS-DFT minimum energy.  This inset demonstrates relatively small differences in the total energies computed with the various embedding and open-shell treatments.

Finally, the dashed black curve in Fig.~\ref{fig:Fe_fig}(a) demonstrates the importance of including spin-dependence in the embedding potential.  This curve corresponds to the RO-DFT-in-DFT-CS treatment of the high-spin state of the hexaaquairon(II) cation described in Sec. III A 1.  It exhibits large relative errors (over 70 kcal/mol) compared to the other treatments of the high-spin state of the hexaaquairon(II) cation, as well as qualitatively incorrectly shortening of the equilibrium internuclear distance.
 Although this approximation is expected to be more reliable for systems in which the spin-density is strongly localized with a single subsystem, the result demonstrates that substantial errors can emerge due to the neglect of spin-dependence in the embedding potential.

\subsubsection{WFT-in-DFT Embedding}

We now consider WFT-in-DFT embedding for the hexaaquairon(II) cation, employing the same subsystem partitioning as in the 
DFT-in-DFT embedding calculations (Fig.~\ref{fig:Fe_fig}(b)).  
The hexaaquairon(II) cation is a benchmark system for spin splittings in transition metal complexes.\cite{Nes04}   
We initially discuss results for MP2 embedding to compare the U-WFT-in-DFT and RO-WFT-in-DFT approaches, and we then present results obtained using CCSD(T) embedding.
 
\begin{figure}[h!] 
  \begin{center}
    \includegraphics[angle=0,width=8.5cm]{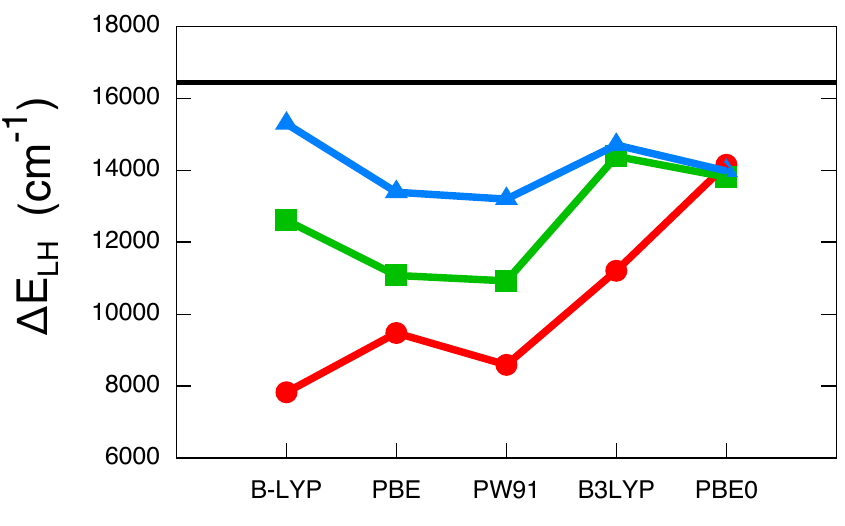}
  \end{center}
  \vspace{-20pt}
\caption{  
MP2-in-DFT embedding for the hexaaquairon(II) cation.  High-spin/low-spin splitting energies obtained using KS-DFT (red,circles), U-MP2-in-DFT (green, squares), and RO-MP2-in-DFT (blue,triangles) with a range of different XC functionals that include B-LYP,\cite{Becke1988,Par97} PBE,\cite{Ernzerhof96,Ernzerhof97} PW91,\cite{Perdew92} B3LYP,\cite{Becke1993} and PBE0.\cite{Barone99} The black line indicates the reference value of 16439 cm$^{-1}$ obtained at the RO-MP2 level of theory; U-MP2 yields a value of 17396 cm$^{-1}$.
}
\label{fig:mp2}
\end{figure}

 Fig.~\ref{fig:mp2} presents results for the low-spin/high-spin energy difference ($\Delta E_{\textrm{LH}}$) obtained using MP2, KS-DFT, and MP2-in-DFT embedding; detailed values are reported in Table~\ref{table:HSLS_MP2}.   
 For KS-DFT calculations of $\Delta E_{\textrm{LH}}$, the energy for the high-spin state of the hexaaquairon(II) cation was obtained at the UKS-DFT level of theory.  The WFT-in-DFT embedding energy for the low-spin state of the hexaaquairon(II) cation is obtained using closed-shell WFT-in-DFT (Sec. II B), 
 while the high-spin state is treated using either U-WFT-in-DFT or RO-WFT-in-DFT (Sec. III A 2). The KS-DFT results (red in Fig.~\ref{fig:mp2}) exhibit strong dependence on the XC functional, with hybrid functionals underestimating $\Delta E_{\textrm{LH}}$ to a somewhat lesser degree than the semi-local functionals.

Fig.~\ref{fig:mp2} clearly illustrates that the RO-MP2-in-DFT results (blue) are in better agreement with the full MP2 calculation than the corresponding U-MP2-in-DFT results (green), particularly for semi-local XC functionals.  
Removal of spin-contamination in the WFT calculation reduces the energy of the high-spin state RO-WFT-in-DFT calculation with respect to that obtained using U-WFT-in-DFT.

Another important observation from Fig.~\ref{fig:mp2} is that the dependence of $\Delta E_{\textrm{LH}}$ on the DFT XC functional is greatly reduced in the embedding calculation, even though only the single transition metal atom is treated at the WFT level.  
 The spread of values obtained at the KS-DFT level of theory is over 6000 cm$^{-1}$, which is reduced by a factor of 3 in the RO-MP2-in-DFT embedding calculations.

\begin{table}
	\caption{
	High-spin/low-spin splitting energies in cm$^{-1}$ for the hexaaquairon(II) cation obtained using KS-DFT, U-MP2-in-DFT, and RO-MP2-in-DFT with a range of different XC functionals.
	\symbolfootnote[0]{$^\mathrm{a}$RO-MP2 yields 16439 cm$^{-1}$.}\symbolfootnote[0]{$^\mathrm{b}$U-MP2 yields   17396 cm$^{-1}$.}$^\mathrm{a,b}$  } 

\begin{tabular}{ | l  | c  | c | c | }
	\hline
Functional  & KS-DFT & U-MP2-in-DFT & RO-MP2-in-DFT \\   \hline
B-LYP  & 7828  & 12604 & 15294  \\ 
PBE  & 9479 & 11079 & 13395  \\ 
PW91  & 8593  & 10924 & 13201 \\ 
B3LYP  & 11206 & 14387 & 14703 \\ 
PBE0  & 14154 & 13812 & 13979   \\ \hline
 \end{tabular}
  \label{table:HSLS_MP2}
 \end{table}

\begin{figure}[h!] 
  \begin{center}
  \hspace*{-20pt}  
    \includegraphics[angle=0,width=8.5cm]{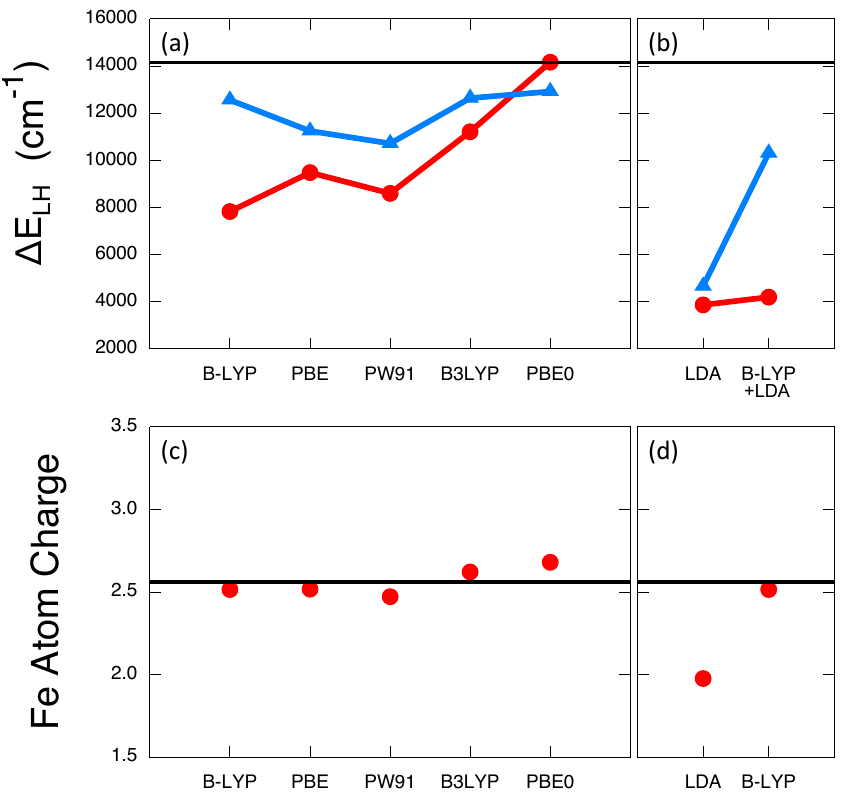}
  \end{center}
  \vspace{-20pt}
\caption{
CCSD(T)-in-DFT embedding for the hexaaquairon(II) cation.  
 (a,b)  High-spin/low-spin splitting energies obtained using KS-DFT (red,circles) and RO-CCSD(T)-in-DFT (blue,triangles) with a range of different XC functionals. The  B-LYP+LDA result is obtained using the B-LYP XC functional for the density calculation and the LDA XC functional for the energy calculation, as is described in the  text. 
 The black line indicates the reference value of 14149 cm$^{-1}$ obtained at the RO-CCSD(T) level of theory. 
(c,d) The charge on the Fe atom is obtained using the Mulliken population analysis of the KS-DFT calculation with each functional.  The relaxed CCSD density, indicated by the black line, has an Fe atomic charge of 2.56.
}
\label{fig:ccsdt}
\end{figure}

Fig.~\ref{fig:ccsdt}(a) presents calculations of the low-spin/high-spin splitting obtained using WFT-in-DFT calculations at the RO-CCSD(T)-in-DFT level of theory; detailed values are reported in Table~\ref{table:HSLS_CCSDT}.  
For the reference calculation obtained at the full RO-CCSD(T) level of theory,\cite{Werner93} no T2 amplitudes were found to exceed 0.05, indicating that a single-reference description of the wavefunction is adequate.
%
The general trend for the RO-CCSD(T)-in-DFT calculations is consistent with the results obtained from RO-MP2-in-DFT.   It is again seen that the dependence of $\Delta E_{\textrm{LH}}$ on the XC functional is substantially reduced using RO-CCSD(T)-in-DFT embedding, and the accuracy of the KS-DFT results are generally improved by treating the transition metal atom at the WFT level.  For this system, the embedded RO-CCSD(T) calculation involves correlating significantly fewer electrons than the full RO-CCSD(T) calculation, and we found that the WFT step in the RO-CCSD(T)-in-DFT calculation required approximately 50 times less wall-clock time than the full RO-CCSD(T) calculation.  

Fig.~\ref{fig:ccsdt}(b) shows that the LDA functional\cite{Sla51,Nus80} presents an interesting outlier compared to the other results in Fig.~\ref{fig:ccsdt}(a). 
Unlike the semi-local and hybrid functionals, RO-CCSD(T)-in-LDA calculations do not exhibit a significant improvement with respect to the corresponding KS-DFT result.  We now show that this anomalous result arises from a density-based error in the LDA functional. 

Fig.~\ref{fig:ccsdt}(c) and Fig.~\ref{fig:ccsdt}(d) present the charge on the Fe atom from a Mulliken population analysis for the low-spin state of the hexaaquairon(II) cation.  Fig.~\ref{fig:ccsdt}(c) shows that the semi-local and hybrid functionals all yield a similar charge for the Fe atom, which is very close to that of the full (relaxed) CCSD density.  In contrast, Fig.~\ref{fig:ccsdt}(d) reveals the LDA functional significantly underestimates the Fe atomic charge, which indicates a significant error in the calculation of the ground state density. Although the use of embedded WFT can be expected to overcome the error in the contribution to the spin-splitting energy due to the LDA functional, it can not overcome this error in the actual ground state density due to LDA.

To confirm this interpretation, we  show that 
removing the error in the LDA density leads to improved WFT-in-DFT estimates for the spin-splitting energy, even if the  LDA functional is still employed for the DFT contributions to the energy.  In Fig.~\ref{fig:ccsdt}(b), the B-LYP+LDA result for WFT-in-DFT embedding  (blue, triangle) is obtained by \emph{(i)} calculating the embedding potential and the subsystem densities using the B-LYP XC functional,  
\emph{(ii)}  performing the embedded WFT calculation at the CCSD(T) level, and 
\emph{(iii)}  using the LDA functional and CCSD(T) to evaluate the respective DFT and WFT contributions to the total energy in Eq.~\ref{osenergy}.
The corresponding B-LYP+LDA result for KS-DFT   (red, circle) is obtained by calculating the total density using KS-DFT with the B-LYP XC functional and then using the LDA functional to evaluate the KS-DFT energy.
As is seen in Fig.~\ref{fig:ccsdt}(d), the B-LYP treatment of the subsystem densities leads to the expected partial charge for the Fe atom; it avoids the error in the electronic density that is introduced using LDA.
However, the  spin-splitting energy obtained using the B-LYP+LDA result for KS-DFT is essentially no better than that  obtained using KS-DFT with the LDA functional (Fig.\ref{fig:ccsdt}(b)), indicating that simply correcting the LDA error in the  density is not enough to avoid the LDA error in the energies.
 Finally, Fig.~\ref{fig:ccsdt}(b) shows that the B-LYP+LDA result for WFT-in-DFT  does exhibit a substantial improvement over the corresponding KS-DFT result; this confirms that  WFT embedding  is able to overcome energy-based errors due to the DFT XC functional, although it is less effective at overcoming density-based errors due to the DFT XC functional.

\begin{table}
	\caption{
		High-spin/low-spin splitting energies  in cm$^{-1}$ for the hexaaquairon(II) cation obtained using KS-DFT and RO-CCSD(T)-in-DFT with a range of different XC functionals.
\footnote{RO-CCSD(T) yields   14149 cm$^{-1}$.}}
\begin{tabular}{ | l  | c  | c |  }
	\hline
Functional  & KS-DFT & RO-CCSD(T)-in-DFT \\ \hline
B-LYP  & 7828  & 12554  \\ 
PBE  & 9479 & 11238  \\ 
PW91  & 8593  & 10712 \\ 
B3LYP  & 11206 & 12634 \\ 
PBE0  & 14154 & 12912   \\ \hline
 \end{tabular}
  \label{table:HSLS_CCSDT}
 \end{table}

Although we have shown that  WFT-in-DFT embedding with the subsystem partitioning shown in Fig.~\ref{fig:Fe_fig}(b) generally leads to improved estimates for the low-spin/high-spin splitting energy over KS-DFT, the same does not hold true for calculated ligation energies of the 
 hexaaquairon(II) cation.  
Ligation energies calculated  using RO-CCSD(T)-in-DFT embedding are essentially unchanged from those obtained  using KS-DFT with the corresponding XC functional; indeed, the mean absolute difference between the computed WFT-in-DFT and KS-DFT ligation energy is only 0.6 kcal/mol per ligand across the set of functionals that includes LDA, B-LYP, PBE, PW91, B3LYP, and PBE0. 
Unlike the spin-splitting energy, which is highly sensitive to the electronic structure of the Fe atom and is thus impacted by the WFT subsystem description, the  ligation energy is dominated by interactions between the Fe atom and the water ligands; these inter-subsystem interactions are still treated essentially at the DFT level  in WFT-in-DFT embedding. An improved description for the ligation energy could be obtained by simply expanding the number of electrons that are treated at the WFT level of theory, or by including two-body correlation corrections through an embedded many-body expansion description of the system.\cite{tfm12jason}

\section*{CONCLUSIONS}

In this work, we have introduced and demonstrated improved methods for the implementation of WFT-in-DFT calculations for open-shell systems and systems with low-lying virtual orbitals.
A simple orbital-occupation-freezing technique is introduced to enable robust 
OEP calculations on 
systems with small HOMO-LUMO gaps, leading to accurate DFT-in-DFT and WFT-in-DFT embedding calculations on transition-metal complexes.
Furthemore, the use of spin-dependent embedding potentials is shown to preserve the accuracy of open-shell DFT-in-DFT calculations in both the restricted and unrestricted orbital formulations, whereas neglect of the spin polarization leads to significant errors in both computed energies and geometries.   
  WFT-in-DFT calculations on the hexaaquairon(II) cation reveal that the treatment of only the single transition metal atom leads to significant improvements in the accuracy of calculated spin-splittings, as well as marked reduction in the dependence of results on the DFT XC functional.
  Taken together, the exact embedding techniques reported and demonstrated here offer a promising approach to the robust treatment of systems for which the accuracy of WFT  is required but for which the cost of the full WFT calculation is not feasible.

\section*{ACKNOWLEDGEMENTS}
This work is supported in part by the Air Force Office of Scientific Research (FA9550-11-1-0288) and the U. S. Army Research Laboratory and the U. S. Army Research Office (W911NF-10-1-0202).
TFM and FRM also gratefully acknowledge network funding from the NSF (CHE-1057112) and EPSRC (EP/J012742/1), respectively. 
Computational
resources were provided by the National Energy Research
Scientific Computing Center, which is supported by the
Office of Science of the US Department of Energy under
Contract No. DE-AC02-05CH11231.

\end{document}